\documentclass[
amssymb, amsmath,%
aip, jmp,
]{revtex4}
\usepackage{graphics}
\usepackage{amssymb,amsmath}
\usepackage[dvips]{lscape,graphicx}

\newcommand{\ct}{\cite}
\newcommand{\lb}{\label}

\newcommand{\bc}{\begin{center}}
\newcommand{\ec}{\end{center}}
\newcommand{\bd}{\begin{displaymath}}
\newcommand{\ed}{\end{displaymath}}
\newcommand{\be}{\begin{equation}}
\newcommand{\ee}{\end{equation}}
\newcommand{\ba}{\begin{array}}
\newcommand{\ea}{\end{array}}
\newcommand{\bea}{\begin{eqnarray}}
\newcommand{\eea}{\end{eqnarray}}
\newcommand{\bt}{\begin{tabular}}
\newcommand{\et}{\end{tabular}}

\newcommand{\bp}{\begin{picture}}
\newcommand{\ep}{\end{picture}}
\newcommand{\bfi}{\begin{figure}}
\newcommand{\efi}{\end{figure}}

\def\fun#1#2{\lower3.6pt\vbox{\baselineskip0pt\lineskip.9pt
\ialign{$\mathsurround=0pt#1\hfil##\hfil$\crcr#2\crcr\sim\crcr}}}

\begin{document}




\title{Gravity and Mirror Gravity in Plebanski Formulation}
\author{D. L.~Bennett}
\email{dlbennett99@gmail.com}
\affiliation{Brookes Institute for Advanced Studies, B{\o}gevej 6, 2900 Hellerup, Denmark}

\author{L. V.~Laperashvili}
\email{laper@itep.ru}
\affiliation{The Institute of Theoretical and
Experimental Physics, Bolshaya Cheremushkinskaya, 25, 117218 Moscow, Russia}

\author{H. B.~Nielsen}
\email{hbech@nbi.dk}
\affiliation{The Niels Bohr Institute, Blegdamsvej 17-21, DK-2100 Copenhagen,
Denmark}

\author{A.~Tureanu}
\email{anca.tureanu@helsinki.fi}
\affiliation {Department of Physics, University of Helsinki, P.O. Box 64, FIN-00014 Helsinki, Finland}



\begin{abstract}
We present several theories of four-dimensional gravity in the
Plebanski formulation, in which the tetrads and the connections
are the independent dynamical variables. We consider the relation
between different versions of gravitational theories: Einstenian,
dual, 'mirror' gravities and gravity with torsion. According to
Plebanski's assumption, our world, in which we live, is described
by the self-dual left-handed gravity. We propose that if the
Mirror World exists in Nature, then the 'mirror gravity' is the
right-handed anti-self-dual gravity with broken mirror parity.
Considering a special version of the Riemann--Cartan space-time,
which has torsion as additional geometric property, we have shown
that in the Plebanski formulation the ordinary and dual sectors of
gravity, as well as the gravity with torsion, are equivalent. In
this context, we  have also developed a 'pure connection gravity'
-- a diffeomorphism-invariant gauge theory of gravity. We have
calculated the partition function and the effective Lagrangian of
this four-dimensional gravity and have investigated the limit of
this theory at small distances.

\end{abstract}


\maketitle

\section{Introduction. Plebanski's formulation of gravity}\label{sec1}

The main idea of Plebanski's formulation of the 4-dimensional
theory of gravity \ct{1} is the construction of the gravitational
action  from the product of two 2-forms \ct{1,2,3,4,5,5a,6}. These
2-forms are constructed using the connection $A^{IJ}$ and tetrads
$\theta^I$ as independent dynamical variables.

We consider a Lorentzian metric. The signature of the metric tensor
denoted by the pair of integers $(p,q)$ is chosen as the Lorentzian
signature $(1,3)$.

The tetrads $\theta^I$ are used instead of the metric
$g_{\mu\nu}$. Both $A^{IJ}$ and $\theta^I$ are 1-forms:
\be   A^{IJ} = A_{\mu}^{IJ}dx^{\mu}  \quad {\mbox{and}}\quad
       \theta^I = \theta_{\mu}^Idx^{\mu}.
                          \lb{1} \ee
Here the indices $I,J = 0,1,2,3$ refer to the space-time with
Minkowski metric $\eta_{IJ}$: $\eta^{IJ} = {\rm diag}(1, -1, -1,
-1)$. This is a flat space which is tangential to the curved space
with the metric $g_{\mu\nu}$. The world interval is represented as
$ds^2 =  \eta_{IJ}\theta^I \otimes \theta^J$, i.e.
\be g_{\mu\nu} = \eta_{IJ} \theta^I_{\mu}\otimes \theta^J_{\nu}.
 \lb{2} \ee

Considering the case of the Minkowski flat space-time with the
group of symmetry $SO(1,3)$, we have the capital latin indices
$I,J,...=0,1,2,3$, which are vector indices under the rotation
group $SO(1,3)$.

In the well-known Plebanski's BF-theory of general relativity (GR)
\ct{1}, the gravitational action (with zero cosmological constant)
is
\be
           \int \epsilon^{IJKL} B^{IJ}\wedge F^{KL},
                    \lb{3} \ee
where $B^{IJ}$ and $F^{IJ}$ are the following 2-forms:
\be
      B^{IJ} = \theta^I\wedge \theta^J = \frac 12
      \theta_{\mu}^I\theta_{\nu}^Jdx^{\mu}\wedge dx^{\nu} ,
                    \lb{4} \ee
and
\be
      F^{IJ} = \frac 12 F_{\mu\nu}^{IJ}dx^{\mu}\wedge dx^{\nu}.
                    \lb{5} \ee
Here the tensor $F_{\mu\nu}^{IJ}$ is the field strength of the
spin connection $A_{\mu}^{IJ}$:
\be
     F_{\mu\nu}^{IJ} = \partial_{\mu}A_{\nu}^{IJ} -
         \partial_{\nu}A_{\mu}^{IJ} - [A_{\mu}, A_{\nu}]^{IJ},
                               \lb{6} \ee
which determines the Riemann--Cartan curvature:
\be
       R_{\mu\nu\lambda\kappa} =  F_{\mu\nu}^{IJ}
 \theta_{\lambda}^I \theta_{\kappa}^J.
                  \lb{7} \ee

Now the question is how many different products of two simple
2-forms can be constructed in the space-time of the 4-dimensional
GR. Then all of these 4-forms can be considered as terms of the
integrand of the gravitational action.

We have only a few possibilities for such 4-forms:
\be B^{IJ}\wedge B^{IJ}, \quad B^{IJ}\wedge F^{IJ}, \quad
F^{IJ}\wedge F^{IJ},  \lb{8} \ee
their dual counterparts:
\be \epsilon^{IJKL}B^{IJ}\wedge B^{KL}, \quad
\epsilon^{IJKL}B^{IJ}\wedge F^{KL}, \quad
\epsilon^{IJKL}F^{IJ}\wedge F^{KL},
 \lb{9} \ee
and
\be S^I\wedge S^I. \lb{10} \ee
The 2-form
\be  S^I =\frac 12 S_{\mu\nu}^Idx^{\mu}\wedge dx^{\nu}
            \lb{31} \ee
contains the torsion $S_{\mu\nu}^I$ \ct{7}:
\be
     S_{\mu\nu}^I = D_{\mu}^{IJ}\theta_{\nu}^J -
     D_{\nu}^{IJ}\theta_{\mu}^J,
                                  \lb{11} \ee
where
\be   D_{\mu}^{IJ} = \delta^{IJ}\partial_{\mu} -  A_{\mu}^{IJ}
                             \lb{12} \ee
is the covariant derivative.

In the Plebanski BF-theory, the gravitational action with nonzero
cosmological constant $\Lambda$ is presented by the integral:
\be  I_{GR} = \frac{1}{\kappa^2}\int \epsilon^{IJKL}\left(B^{IJ}\wedge
    F^{KL} + \frac{\Lambda}{4}B^{IJ}\wedge B^{KL}\right),
                                  \lb{13} \ee
where $\kappa^2=8\pi G$, $G$ is the gravitational constant, and
the reduced Planck mass is $M_{Pl}^{red.} = 1/{\sqrt{8\pi G}}$.

Below we use units $\kappa=1$ (in these units $M_{Pl}^{red.}=1$).

The "dual sector" of gravity is described by the following
integral \ct{8}:
\be  I_{dual\,\,GR} = 2 \int\left (B^{IJ}\wedge
    F^{IJ} + \frac{\Lambda}{4}B^{IJ}\wedge B^{IJ}\right) =
    \int \epsilon^{IJKL}\left(B^{IJ}\wedge F^{*KL}
    + \frac{\Lambda}{4}B^{IJ}\wedge B^{*KL}\right),
                                  \lb{14} \ee
where $F^{*IJ} = \frac 12 \epsilon^{IJKL} F^{KL}$ is the dual
tensor.

The paper is organized as follows. In Section \ref{sec1} we review  the
main idea of Plebanski to construct the 4-dimensional theory of
gravity
considering the integrand of the gravitational action as the
product of two 2-forms containing only tetrads and connections
which are independent dynamical variables. The BF-theories are
presented for the ordinary and dual gravity in the Minkowski flat
space-time. In Section \ref{sec2} we construct the self-dual left-handed
and the anti-self-dual right-handed gravitational worlds. In
Section \ref{sec3} we investigate the 'mirror gravity' existing in the 'Mirror
World', which describes the states of particle physics with
opposite chirality. We assume that the mirror gravity which has to
interact with the states of opposite "right-handed" chirality  can
be described by the anti-self-dual right-handed action of gravity.
Using modern astrophysical and cosmological measurements, we
consider the broken mirror parity (MP), and discuss the
communications between visible and hidden worlds. Section \ref{sec4} is
devoted to  gravity with torsion. It was shown that in the
self-dual Plebanski formulation of GR, gravity with torsion
coincides with the ordinary and dual versions of gravity.
A new type of gauge transformation in Riemann--Cartan space-time
is considered: Einstein's theory of gravity described only by
curvature can be rewritten as Einstein's teleparallel theory of
gravity described only by torsion. In Section \ref{sec5} the equations of
motion resulting from Plebanski's gravitational action are used to
construct the action containing only the connection and auxiliary
fields. Einstein's equations are investigated in terms of
Plebanski's theory of gravity. Section \ref{sec6} is devoted to the
diffeomorphism invariant gauge theory of gravity, which is a new
type of gravitational theory with a 'pure connection' formulation
of GR. The calculation of the partition function and the effective
Lagrangian of this 4-dimensional gravity is presented, as well as
field equations. The limits of this theory at small and large
distances are investigated. In the asymptotic limit of this
theory, we have gravity in the flat (Euclidean or Minkowski)
space-time and the effective gravitational coupling constant is
given by the bare cosmological constant $\Lambda$:
$g_{eff}=\Lambda/2$. At large distances we predict a more
complicated theory of gravity. The perspectives of the quantum
theory of Plebanski's gravity are discussed in Subsections \ref{sec6.1} and
\ref{sec6.2}. Section \ref{sec7} contains a summary and conclusions.

\section{The "left-handed" and "right-handed" gravity}\label{sec2}

The next step of the study of the Plebanski BF-theory is the
consideration of the decomposition of $SO(1,3)$-group of GR on
left- and right-handed sectors.

For any antisymmetric tensor $A^{IJ}$ there exists a dual tensor
given by the Hodge star dual operation on the indexes $I,J$ of
flat space:
\be
           A^{*IJ} = \frac 12 \epsilon^{IJKL}A^{KL}.
                                 \lb{39} \ee
This antisymmetric tensor $A^{IJ}$ can be split into a self-dual
component $A^+$ and an anti-self-dual component $A^-$, according
to the relation:
\be
           A^{\pm} = \frac 12 (A \pm A^*).
                                 \lb{40} \ee

As with any Lie group, the best way to study many aspects of the
Lorentz group is via its Lie algebra. Since the Lorentz group is
$SO(1,3)$, its Lie algebra is reducible and can be decomposed into two
copies of the Lie algebra of $SL(2,R)$:
\be
            SO(1,3) = SL(2,R)_{left}\times SL(2,R)_{right}.
                                 \lb{40a} \ee
As it was shown explicitly in \ct{8a}, this is the Minkowski space
analog of the $SO(4)= SU(2)_{left}\times SU(2)_{right}$
decomposition in a Euclidean space.

The complex Lorentz algebra splits into two complex $SO(3)$
algebra called the self-dual (left-handed) and anti-self-dual
(right-handed) components \ct{1,2}:
\be
            SO(1,3,C) = SO(3,C)_{left}\times SO(3,C)_{right}.
                                 \lb{40b} \ee
Because of this split, the curvature of the self-dual components of
the connection is the self-dual component of the curvature.

In particle physics, a state that is invariant under one of these
copies of $SO(3,C)$ is said to have chirality, and is either
left-handed or right-handed, according to which copy of $SO(3,C)$
it is invariant under.

Self-dual tensors transform non-trivially only under
$SO(3,C)_{left}$ and are invariant under $SO(3,C)_{right}$. By
this reason, they are called "left-handed" tensors. Similarly,
anti-self-dual tensors, non-trivially transforming only under
$SO(3,C)_{right}$, are called "right-handed" tensors. These
self-dual and anti-self-dual tensors $A^{\pm IJ}$ have only three
independent components given by $IJ=0i,\,\,i=1,2,3$:
\be
                 A^{\pm i} = \pm 2A^{\pm 0i},
                     \lb{42} \ee
which transform as adjoint vector components under the
corresponding $SU(2)$-group.

Such a decomposition shows (see Refs.~\ct{2,3,4,5}) that the
actions $I_{GR}$ and $I_{dual\,\,GR}$, given by Eqs.~(\ref{13})
and (\ref{14}), respectively, can be represented in terms of the
"left-handed" and "right-handed" gravity:
\be  I_{GR} =  \int [\Sigma^i\wedge F^i - {\bar{\Sigma}}^i\wedge
{\bar F}^i + \Lambda(\Sigma^i\wedge \Sigma^i -
{\bar{\Sigma}}^i\wedge {\bar{\Sigma^i}})],
                      \lb{43} \ee
and
\be  I_{dual\,\,GR} =  \int [\Sigma^i\wedge F^i +
{\bar{\Sigma}}^i\wedge {\bar F}^i + \Lambda(\Sigma^i\wedge
\Sigma^i + {\bar{\Sigma}}^i\wedge {\bar{\Sigma^i}})],
                      \lb{44} \ee
where
\be F^{\pm i}_{\mu\nu} =
\partial_{\mu}A^{\pm i}_{\nu} -
\partial_{\nu}A^{\pm i}_{\mu} + \epsilon^{ijk}A^{\pm j}_{\mu}A^{\pm k}_{\nu},
\lb{44a} \ee
 $F\equiv F^+,\quad \bar F\equiv F^-,\quad \Sigma\equiv
\Sigma^+, \quad \bar{\Sigma}\equiv
 \Sigma^-,$ and the left-handed and right-handed $\Sigma^{\pm i}$ are given by:
\be
     \Sigma^{+} = i\theta^0\wedge \theta^i - \frac 12
     \epsilon^{ijk}\theta^j\wedge \theta^k,
     \qquad \Sigma^{-} = i\theta^0\wedge \theta^i + \frac 12
     \epsilon^{ijk}\theta^j\wedge \theta^k.
                          \lb{45} \ee
The choice of the second Killing form in the cosmological term is
made so that the full Plebanski action, obtained by adding the
so-called simplicity constraints to \eqref{43}, is equivalent (in
the appropriate sector) to ordinary gravity (see also \ct{5a}).
The so-called simplicity constraints are introduced in the
gravitational actions by means of the Lagrange multipliers
$\psi_{ij}$, which are considered in theory as auxiliary fields,
symmetric and traceless. Finally, the resulting actions of the
Plebanski gravity are \ct{1,2,3,4,5,5a,6}:
\be  I_{GR} =  \int [\Sigma^i\wedge F^i - {\bar{\Sigma}}^i\wedge
{\bar F}^i + (\Psi^{-1})_{ij}(\Sigma^i\wedge \Sigma^j -
{\bar{\Sigma}}^i\wedge {\bar{\Sigma^j}})],
                      \lb{46} \ee
and
\be  I_{dual\,\,GR} = \int [\Sigma^i\wedge F^i +
{\bar{\Sigma}}^i\wedge {\bar F}^i +
(\Psi^{-1})_{ij}(\Sigma^i\wedge \Sigma^j + {\bar{\Sigma}}^i\wedge
{\bar{\Sigma^j}})],
                      \lb{47} \ee
where
\be   (\Psi^{-1})_{ij} = \Lambda\delta_{ij} + \psi_{ij}.
                     \lb{48} \ee
The stationarity with respect to $\psi_{ij}$ provides the correct
number of constraints, reducing the 36 degrees of freedom of
($\Sigma^+,\Sigma^-$) to the 16 degrees of freedom of tetrads
$\theta_{\mu}^I$.

Now we can distinguish the two worlds -- two sectors of gravity:
left-handed gravity and right-handed gravity. The self-dual
left-handed gravitational world can be described by the action:
\be
      I_{(left\,\, GR)}(\Sigma,A) =  \int [\Sigma^i\wedge F^i +
 (\Psi^{-1})_{ij}\Sigma^i\wedge \Sigma^j] ,
                      \lb{52} \ee
while the anti-self-dual right-handed gravitational world is given
by the following action:
\be
 I_{(right\,\, GR)}(\bar \Sigma, \bar A) = \int [{\bar{\Sigma}}^i\wedge {\bar F}^i
 + (\Psi^{-1})_{ij}{\bar{\Sigma}}^i\wedge {\bar{\Sigma^j}}].
                      \lb{53} \ee
If the anti-self-dual right-handed gravitational world is absent
in Nature ($\bar F=0$ and $\bar \Sigma = 0$), then gravity is
presented only by the self-dual left-handed Plebanski action
(\ref{52}). The main assumption of Plebanski was that our world,
in which we live, is a self-dual left-handed gravitational world
described by the action (\ref{52}). The same self-dual formulation
of general relativity (GR) was developed later by Ashtekar \ct{9}.

If there are exist in Nature two parallel worlds with opposite
chiralities -- Ordinary and Mirror (see below, Section \ref{sec3}) -- then we
must consider the left-handed gravity in the Ordinary world and
the right-handed gravity in the Mirror world.

It is not difficult to show \ct{1} that the Plebanski action
(\ref{52}) corresponds to the Einstein--Hilbert action of gravity.
In the Minkowski space background, the Einstein--Hilbert action is:
\be I_{EH} = \frac{1}{\kappa^2}\int \left(\frac 12 R -
\Lambda_0\right)\sqrt{-g}d^4x, \lb{62} \ee
where $R$ is a scalar curvature, and $\Lambda_0$ is Einstein's
cosmological constant. Here we have (see  Subsection \ref{sec5.1}):
\be \Lambda_0 = 6\Lambda. \lb{63} \ee
According to Eqs.~(\ref{46}) and (\ref{47}), in the Plebanski
self-dual formulation of gravity we have the following equality:
\be     I_{GR} = I_{dual\,\,GR}.
 \lb{54} \ee
Both actions are given by the formula (\ref{52}).

\section{Mirror world and Mirror gravity}\label{sec3}

Previously, in Refs.~\ct{10,11} we have assumed that there exists
in Nature a Mirror World (MW) \ct{12,13}, which is a duplication
of our Ordinary World (OW), or shadow Hidden World (HW)
\ct{13a,13b}, parallel to our Ordinary World (OW). This MW (or HW)
can explain the origin of dark matter and dark energy.

Postulating the existence of the Mirror World, we confront
ourselves with a question: should the mirror gravity be the
anti-self-dual right-handed gravity? We assume that we have this
case, and the mirror gravitational action is given by
Eq.~(\ref{53}) describing the anti-self-dual right-handed gravity.

The MW is a mirror copy of the OW and contains the same particles
and types of interactions as our visible world. Lee and Yang were
the first \ct{12} to suggest such a duplication of the worlds,
which restores the left-right symmetry of Nature. The term 'Mirror
Matter' was introduced by Kobzarev, Okun and Pomeranchuk \ct{13}.
They suggested the 'Mirror World' as the hidden sector of our
Universe, which interacts with the ordinary (visible) world only
via gravity or another very weak interaction. This assumption was
considered in many papers.

Considering only pure gravity, we can formulate mirror parity (MP)
investigating the invariance of the Plebanski gravitational action
under the dual symmetry, i.e. under the interchanges:
\be A \leftrightarrow \bar A \quad (F \leftrightarrow \bar F),
\quad \Sigma \leftrightarrow \bar{\Sigma}.
  \lb{63a} \ee
Introducing projectors on the spaces of the so-called self- and
anti-self-dual tensors, we obtain:
\be  P^{\pm} = \frac 12 \left({\cal I}^{IJKL} \pm \frac 12
\epsilon^{IJKL}\right),
  \lb{63b} \ee
where
\be       {\cal I}^{IJKL} = \frac 12 (\delta_{IK}\delta_{JL} -
\delta_{IL}\delta_{JK}).
  \lb{63c} \ee
If we represent the gravitational action (\ref{13}) as
\be
        I = \int \epsilon^{IJKL}{\cal L}^{IJKL},
          \lb{63d} \ee
we can consider the relation:
\be
       P^{+}{\cal L}^{IJKL} =\alpha P^{-} {\cal L}^{IJKL}.
          \lb{63e} \ee
The dual symmetry gives $\alpha=1$. In this case the action
\be
      I(\Sigma,A) =  \int [\Sigma^i\wedge F^i +
 (\Psi^{-1})_{ij}\Sigma^i\wedge \Sigma^j]
                      \lb{83f} \ee
is invariant under the interchanges (\ref{63a}). Considering the
left-handed gravitational action (\ref{52}) in the OW and the
right-handed gravitational action (\ref{53}) in the MW, we have an
unbroken mirror parity (MP). In this case, the bare cosmological
constants in the OW and MW are identical:
\be
             \Lambda_0 =  \Lambda_0^{(O)}  = \Lambda_0^{(M)}.
                                      \lb{83g} \ee

Let us consider now the Universe with matter fields.  Assuming the
existence of the mirror world, we can enlarge the Standard Model
(SM) gauge group $G_{SM} = SU(3)_c \times SU(2)_L \times U(1)_Y$
to $G_{SM} \times G'_{SM}$, where the gauge group $G'_{SM} =
SU(3)'_c \times SU(2)'_R \times U(1)'_Y$ (see Refs.~\ct{14,14a}
and review \ct{15}) is a mirror of $G_{SM}$ with identical gauge
couplings, under which the matter contents switch their
chiralities. Hence, the mirror parity is restored in the Universe
where the visible and mirror worlds coexist in the same
space-time.

Astrophysical and cosmological observations \ct{16} have revealed
the existence of dark matter ($DM$) which constitutes about 23\%
of the total energy density of the present Universe. This is five
times larger than all the visible matter, $\Omega_{DM}: \Omega_{M}
\simeq 5 : 1$.  In parallel to the visible world, the mirror world
conserves mirror baryon number and thus protects the stability of
the lightest mirror nucleon. Mirror particles have therefore been
suggested as candidates for the inferred dark matter in the
Universe \ct{15a} (see also Refs.~\ct{14,14a,15,16a,16b}. This
explains the right amount of dark matter, which is generated via
the mirror leptogenesis \ct{11,16c}, just like the visible
matter is generated via ordinary leptogenesis \ct{16d}.

If the ordinary and mirror worlds are identical, then O- and
M-particles should have the same cosmological densities. But this
is in conflict with recent astrophysical
measurements \ct{16}. Mirror parity (MP) is not conserved, and the
ordinary and mirror worlds are not identical \ct{10,11,14,14a,15}
in the sense that, although the chain of breakings of the gauge
groups is the same in both worlds, the energy scales at which
these breakings take place are different.

Superstring theory also predicts that there may exist in the
Universe another form of matter -- hidden 'shadow matter', which
only interacts with ordinary matter via gravity or
gravitational-strength interactions \ct{13b}.  According to the
superstring theory, the two worlds, ordinary and shadow, can be
viewed as parallel branes in a higher dimensional space, where
O-particles are localized on one brane and hidden particles -- on
another brane, and gravity propagates in the bulk.

In the presence of matter, the Einstein field equations:
\be R^{\mu\nu} - \frac{1}{2}Rg^{\mu\nu}= 8\pi
GT^{\mu\nu} - \Lambda_0  g^{\mu\nu}    \lb{64a} \ee
contain the energy momentum tensor of matter $T^{\mu\nu}$, and all
quantum fluctuations of the matter contribute to the vacuum energy
$\rho_{vac}$ of the Universe. The resulting cosmological constant
is the effective cosmological constant which is equal to
\be \Lambda_{eff} = \Lambda_0  + {8\pi G}\rho_{vac}.
 \lb{64b} \ee
If the mirror parity is broken also in the gravitational sector
($\alpha\neq 1$), then we can distinguish bare cosmological
constants of the O- and M-worlds:
\be \Lambda_0^{(O)}\neq \Lambda_0^{(M)} \lb{64c}. \ee
In the units $\kappa=\kappa'=1$, we have:
\be \Lambda^{(O,M)}_{eff} = \Lambda_0^{(O,M)}  +
\rho_{vac}^{(O,M)} =\rho_{vac.eff}^{(O,M)}.
 \lb{65} \ee
The vacuum energy densities $\rho_{vac}^{(O,M)}$ are given by a
trace of the stress-energy tensor of matter $T_{\mu\nu}^{(O,M)}$
in the O- and M-worlds. The effective vacuum energy of the
Universe is the sum:
\be \Lambda_{eff} = \Lambda^{(O)}_{eff} + \Lambda^{(M) }_{eff}=
\rho_{vac.eff}.
 \lb{66} \ee
Since the ordinary and mirror worlds are not identical, we have:
\be \Lambda^{(O)}_{eff} \neq \Lambda^{(M)}_{eff}.
 \lb{67} \ee
With the aim to explain the tiny value of the dark energy density
$\rho_{DE} = \Lambda_{eff} = \rho_{vac.eff}\simeq (2.3\times
10^{-3}\,\,{\mbox{eV}})^4$, verified by astronomical and cosmological
observations \ct{16},  we have several possibilities. For example:\\

I) The Universe is described by the theory $G_{SM} \times
G'_{SM}$ with broken mirror parity \ct{14,14a,15}. We can assume
that
   $$ \Lambda^{(O)}_{eff}= \Lambda_0^{(O)}  + \rho_{vac}^{(SM)}
   \simeq 0,\quad
   \Lambda^{(M)}_{eff} = \Lambda_0^{(M)}  + \rho_{vac}^{(SM')}\simeq
   \rho_{DE}.$$
If the supersymmetry breaking scale is the same in O- and
M-worlds, then $\rho_{vac}^{(SM)}\simeq  \rho_{vac}^{(SM')}$ and
$$\rho_{DE}\simeq \Lambda_0^{(M)} - \Lambda_0^{(O)}.$$
Then the condensate of gravitational fields can explain the value of $\rho_{DE}$.\\

II) In Refs.~\ct{10,11} we considered the theory of the $E_6$
unification with different types of breaking in the visible (O)
and hidden (M) worlds. We assumed that at the first stage of the
Universe we had unbroken mirror parity:
$\Lambda_0^{(O)}=\Lambda_0^{(M)}$ and $E_6=E'_6$. Finally, $E_6$
was broken to $G_{SM}$, and $E'_6$ underwent the breaking to
$G'_{SM}\times SU(2)'_{\theta}$, where $SU(2)'_{\theta}$ is the
group whose gauge fields are massless vector particles, 'thetons'
\ct{20a} . These 'thetons' have a macroscopic confinement radius
$1/\Lambda'_{\theta}$. The estimate given by Refs.\ct{10,11}
confirms the scale $\Lambda'_{\theta} \sim 10^{-3}$ eV.

We assumed that $\Lambda^{(O)}_{eff} = \Lambda_0^{(O)}  +
\rho_{vac}^{(SM)} \simeq 0$, and $\Lambda^{(M)}_{eff} =
\Lambda_0^{(M)}  + \rho_{vac}^{(SM')} + \rho_{vac}^{(\theta')}\neq
0$. Assuming the same supersymmetry breaking scale in O- and
M-worlds, we considered: $\Lambda_0^{(M)}  + \rho_{vac}^{(SM')}
\simeq 0$. Then:
\be \Lambda_{eff} = \Lambda^{(M)}_{eff} = \rho_{DE} \simeq
\rho_{vac}^{(\theta')}\simeq {(\Lambda'_{\theta})}^4\simeq
(2.3\times 10^{-3}\,\,{\mbox{eV}})^4, \lb{68} \ee
in accordance with cosmological measurements \ct{16}.

\subsection{Communications between Visible and Hidden Worlds}

Mirror particles have not been seen so far, because the
communication between visible and hidden worlds is hard.

There are several fundamental ways by which the hidden world can
communicate with our visible world.

I) It is necessary to assume that the self-dual gravity interacts
not only with visible matter, but also with mirror matter, and
anti-self-dual gravity also interacts with matter and mirror
matter (see Subsection \ref{sec6.1}).
It is then to be expected that a fraction of the mirror matter
exists in the form of mirror galaxies, mirror stars, mirror
planets etc. These objects can be detected using gravitational
microlensing \ct{20b}.

II) There exists the kinetic mixing between the electromagnetic
field strength tensors for visible and mirror photons:
\be         L_{\gamma}^{{mix}} = - \frac{\epsilon_{\gamma}}{2}
F_{\gamma\,\mu\nu}{F'_{\gamma}}^{\mu\nu}.  \lb{70} \ee
The photon--mirror photon mixing induces oscillations between
orthopositronium and mirror orthopositronium \ct{20c}.
Orthopositronium could then turn into mirror orthopositronium and
then decay into mirror photons.

Besides these interactions, the hidden world can communicate with
our visible world by mass mixings between visible and mirror
neutrinos \ct{20d} (neutrino--mirror neutrino oscillations). Also
mirror neutrons can oscillate to ordinary neutrons \ct{20e}. It is
expected the interaction between visible and mirror quarks,
leptons and Higgs bosons, etc. (see \ct{11,14,15,16b}).
The search for mirror particles at LHC is discussed in Ref.~\ct{20f}.

Heavy Majorana neutrinos $N_a$ are singlets of $G_{SM}$ and
$G'_{SM}$, and they can be messengers between visible and hidden
worlds \ct{11,16c}.

The dynamics of the two worlds of our Universe, visible and hidden, is
governed by the following action:
\be   I = \int [ L_{grav} + L'_{grav} + L_M + L'_M + L_{mix}],
                                                   \lb{71} \ee
 where $L_{grav}$ is the gravitational (left-handed) Lagrangian in
the visible world, and $L'_{grav}$ is the gravitational
right-handed Lagrangian in the hidden world,
$L_M\,(L'_M)$ is the matter Lagrangian in the O(M)-world, and
$L_{mix}$ is the Lagrangian describing all mixing terms (see
\ct{14,15}). Mixing terms give very small contributions to
physical processes.

\section{Torsion in Plebanski's formulation of gravity}\label{sec4}

The gravitational theory with torsion can be presented in the
Plebanski formalism by the following integral:
\be I_S = 2 \int \left(2 S^I\wedge S^I + \frac{\Lambda}{4}B^{IJ}\wedge
B^{IJ}\right).  \lb{15} \ee
Using the partial integration and putting
\be \int \partial_{\mu}(T_{\nu\kappa\lambda})dx^{\mu}\wedge
dx^{\nu}\wedge dx^{\kappa}\wedge dx^{\lambda}=0,
 \lb{16} \ee
it is not difficult to show that
\be \int B^{IJ}\wedge
    F^{IJ} = 2\int S^I\wedge S^I.  \lb{17} \ee
According to Eqs.~(\ref{14}) and (\ref{15}), we have:
\be
 I_{dual\,\,GR} = 2 \int \left(B^{IJ}\wedge
    F^{IJ} + \frac{\Lambda}{4}B^{IJ}\wedge B^{IJ}\right) =
2 \int \left(2 S^I\wedge S^I + \frac{\Lambda}{4}B^{IJ}\wedge B^{IJ}\right) =
I_S. \lb{18} \ee
This means that in the self-dual Plebanski formulation of the
gravitational theory the following sectors of gravity coincide:
\be
  I_{GR} = I_{dual \,\,GR} = I_S.
 \lb{19} \ee

Now it is obvious that we can exclude torsion as a separate
dynamical variable. Here we see a close analogy of the  geometry
of the curved Riemann--Cartan space-time with torsion \ct{7}, with
a structure of defects in a crystal \ct{21,21a,21b}.

A crystal can have two different types of topological defects
\ct{21}. A first type of such defects are translational defects
called {\it dislocations}: a part of a single-atom layer is
removed from the crystal and the remaining atoms relax to
equilibrium under the elastic forces. A second type of defects is
of the rotation type and called {\it disclinations}. They arise by
removing an entire wedge from the crystal and re-gluing the free
surfaces.

The geometry of the 4-dimensional Riemann--Cartan space-time is
described by the direct generalizations of the translational and
rotational defect gauge fields of the 3-dimensional crystal to the
tetrads $\theta_{\mu}^I$ and connections $A_{\mu}^{IJ}$, which
play the role of the translational and rotational defect gauge fields
in the 4-dimensional Riemann--Cartan space-time \ct{21}.

The field strength of $A_{\mu}^{IJ}$ is given by the tensor
(\ref{6}), which describes the space-time curvature, and the field
strength of $\theta_{\mu}^I$ is the torsion given by
Eq.~(\ref{11}). Torsion is presented by dislocations, and
curvature by disclinations. But these defects are not independent
of each other: a dislocation is equivalent to a
disclination-antidisclination pair, and a disclination presents a
string of dislocations. This explains why Einstein's theory of
gravity described only by curvature can be rewritten as
Einstein's "teleparallel"\, theory of gravity \ct{22} described
only by torsion.

In summary, we wish to emphasize that if the Einstein--Hilbert
Lagrangian is expressed in terms of the translational and
rotational gauge fields, the tetrads $\theta^I_{\mu}$ and the
connection $A^{IJ}_{\mu}$, then  the Cartan curvature can be
converted to torsion and back, totally or partially, by a new type
of gauge transformation in Riemann--Cartan space-time \ct{21}.

In this general formulation, Einstein's original theory is
obtained by going to the zero-torsion gauge, while the
"teleparallel"\, theory is obtained in the gauge in which the Cartan
curvature tensor vanishes. But any intermediate choice of the
field $A^{IJ}_{\mu}$ is also allowed.

\section{Equations of motion}\label{sec5}

The equations of motion resulting from Plebanski's action of
gravity given by Eq.~(\ref{52}) (with $I\equiv I_{(left\,\, GR)}$)
are:
\be  \frac{\delta I}{\delta A^i} = D\Sigma^i = d\Sigma^i +
\epsilon^{ijk}A^j \wedge \Sigma^k = 0,    \lb{20} \ee
\be  \frac{\delta I}{\delta \psi_{ij}} = \Sigma^i \wedge \Sigma^j
- \frac 13 \delta^{ij} \Sigma^k \wedge \Sigma^k = 0,    \lb{21}
\ee
\be     \frac{\delta I}{\delta \Sigma^i} = F^i -
(\Psi_{ij})^{-1}\Sigma^j = 0.     \lb{22} \ee
Eq.~(\ref{20}) states that $A^i$ is the self-dual part of the spin
connection compatible with the 2-forms $\Sigma^i$, where $D$ is
the exterior covariant derivative with respect to $A^i$.
Eq.~(\ref{21}) implies that the 2-forms $\Sigma^i$ can be
constructed from tetrad one-forms giving (\ref{45}), which fixes
the conformal class of the space-time metric $g_{\mu\nu} =
\eta_{IJ}\theta^I_{\mu} \otimes \theta^J_{\nu}$ defined by
tetrads. Eq.~(\ref{22}) states that
the trace-free part of the Ricci tensor
vanishes \ct{1,2,4}.

The 2-form fields $\Sigma^i$ can therefore be integrated out of
Eq.~(\ref{52}). Thus, we are led to Plebanski's gravity given by
the form:
\be
    I_{(left\,\, GR)}(A,\psi) =  \int \Psi_{ij}F^i\wedge F^j = \int {(\Lambda\delta_{ij}
    +\psi_{ij})}^{-1}F^i\wedge F^j,  \lb{23} \ee
discussed in Refs.~\ct{1,2,3,4,5,5a,6}, and
\be
    I_{(right\,\, GR)}(\bar A,\psi') =  \int \Psi'_{ij}{\bar F}^i\wedge {\bar F}^j = \int {(\Lambda'\delta_{ij}
    +\psi'_{ij})}^{-1}{\bar F}^i\wedge {\bar F}^j.  \lb{23a} \ee

\subsection{Einstein's equations}\label{sec5.1}

Let us consider Einstein's equations in terms of the
self-dual Plebanski's theory of gravity (we assume O-world).
From Eq.~(\ref{22}) we obtain:
\be  F^i \wedge \Sigma^i = (\Lambda\delta_{ij} +
\psi_{ij})\Sigma^i \wedge \Sigma^j. \lb{24} \ee
Here,
\be  iF^i \wedge \Sigma^i = \frac 14 \epsilon^{\mu\nu\rho\sigma}
F_{\mu\nu}^i\Sigma_{\rho\sigma}^i \sqrt -g d^4x. \lb{25} \ee
According to Ref.~\ct{4}:
\be    i\Sigma^i\wedge \Sigma^j = 2\delta^{ij} \sqrt -g d^4x.
 \lb{59} \ee
Taking into account that ${\rm{Tr}}\,\psi_{ij}=0$, we have:
\be
 i(\Lambda \delta_{ij} + \psi_{ij})\Sigma^i\wedge \Sigma^j =
 2(3\Lambda + {\rm{Tr}}\,\psi_{ij})\sqrt -g d^4x = 6\Lambda \sqrt -g
 d^4x,
 \lb{60} \ee
what gives:
\be \frac 14 \epsilon^{\mu\nu\rho\sigma}
F_{\mu\nu}^i\Sigma_{\rho\sigma}^i =
 6\Lambda = \Lambda_0.
                         \lb{25} \ee
The curvature is a 2-form, and can be split in the basis of
self-dual $\Sigma^i$ and anti-self-dual $\bar{\Sigma}^i$:
\be
       F^i = X^{ij}\Sigma^j + {\bar X}^{ij}{\bar{\Sigma}}^j,
 \lb{26} \ee
where:
\be X^{ij} = \frac 14 \epsilon^{\mu\nu\rho\sigma}
F_{\mu\nu}^i\Sigma_{\rho\sigma}^j.
 \lb{27} \ee
From  Eq.~(\ref{25})) we obtain:
\be
      {\rm{Tr}} X^{ij} = \sum_i X^{ii} = \frac 14
\epsilon^{\mu\nu\rho\sigma}F_{\mu\nu}^i\Sigma_{\rho\sigma}^i.
           \lb{28} \ee
Calculating the matrices $X^{ij}$, we obtain ten equations in
Plebanski's variables, equivalent to the vacuum Einstein's
equations:
\be
   {\rm {Tr}} X^{ij} = \Lambda_0 \quad {\rm{and}} \quad {\bar X}^{ij} =
    0.  \lb{29} \ee
We have the following ten equations with matter fields, equivalent
to Einstein's equations (\ref{64a}) :
\be
   {\rm {Tr}} X^{ij} = \Lambda_0 - 2\pi GT \quad {\rm{and}} \quad {\bar X}^{ij} =
    0.  \lb{30} \ee
Here $T$ is the trace of the stress-energy tensor of matter
$T_{\mu\nu}$.

However, if we have two worlds OW and MW (or HW), then we have two
Einstein's equations:
\be
   {\rm {Tr}} X^{ij} = \Lambda_0^{(O)} - 2\pi GT^{(O)},
     \lb{30a} \ee
and
\be
   {\rm {Tr}}{\bar X}^{ij} = \Lambda_0^{(M)} - 2\pi GT^{(M)},
     \lb{30b} \ee
where
\be
      {\rm{Tr}}{\bar X}^{ij} = \sum_i{\bar X}^{ii} = \frac 14
\epsilon^{\mu\nu\rho\sigma}{\bar F}_{\mu\nu}^i{\bar
\Sigma}_{\rho\sigma}^i.
           \lb{30c} \ee

\subsection{Newtonian gravity}

In the non-relativistic limit we have $A_0^i = \partial_i
\Phi(x)$, where $\Phi(x)$ is given by $g_{00}=1+2\Phi(x)$. Then
\be    \Delta \Phi(x) = 4\pi G\rho, \lb{49r} \ee
what leads to the Newtonian gravitational potential of the
particle with mass $M$ and $G_N=G$:
\be     V (r) \equiv \Phi(r ) \approx - G_N\frac{M}{r}. \lb{50r}
\ee
This result comes from Eq.~(\ref{30}).

\section{Diffeomorphism invariant gauge theory of gravity}\label{sec6}

Gravity is not a gauge theory of the usual type. The carriers of
force in a usual gauge theory are spin one particles. Moreover, in
the electromagnetic field theory, for example, there are two types
of charged objects, negatively and positively charged particles,
which interact by exchange of carriers of force. As a result,
particles can either repel, or attract. In contrast, there is only
one type of charge in gravity, and we have only attraction.

However, scattering amplitudes for gravitons can be expressed as
squares of amplitudes for gluons (see for example \ct{26,27}): the
closed string theory describes gravity, and the open string theory
is a gauge theory. The relationship is not direct, but it exists,
and it is not easy to find a Lagrangian version of the
correspondence. In the Plebanski--Ashtekar formalism the
gravity/gauge theory relationship was developed in
Refs.~\ct{28,29}.

Finally, it was realized that Plebanski's formulation of GR \ct{1}
can be integrated out to obtain a 'pure connection' formulation of
GR, where the only dynamical field is an $SU(2)$ connection
\ct{29,30}. The result is a completely new perspective on general
relativity, in which GR was reformulated as a {\it diffeomorphism
invariant gauge theory}.

The pure connection formulation of GR \ct{29} was further
developed. It was shown in Ref.~\ct{31} that in this case we have
not a single diffeomorphism invariant gauge theory, but an
infinite parameter class of them. All these theories have the same
number of propagating degrees of freedom (DOF). For any theory of
this type the phase space is that of an $SU(2)$ gauge theory.
However, in addition to the usual $SU(2)$ gauge rotations, there
are also diffeomorphisms acting on the phase space variables,
which reduce the number of propagating DOF from 6 of the $SU(2)$
gauge theory to 2 of GR \ct{32}.

In the present paper we try to develop the gravity/gauge theory
correspondence.

Using the imaginary time $x_0=it$, e.g. assuming the Euclideanized
self-dual Plebansky action (\ref{23}), we can calculate the
partition function:
\be    Z = \int[{\cal D}A ][{\cal D}\psi] e^{-S} =
        \int[{\cal D}A ][{\cal D}\psi]{\rm {exp}}\left[-\int {(\Lambda\delta_{ij}
    +\psi_{ij})}^{-1}F^i\wedge F^j\right].  \lb{31} \ee
Here,
\be F^i\wedge F^j = \frac 14 F_{\mu\nu}^i F_{\rho\sigma}^j
dx^{\mu}\wedge dx^{\nu}\wedge dx^{\rho}\wedge dx^{\sigma} = \frac
14 F_{\mu\nu}^i F_{\rho\sigma}^j
\epsilon^{\mu\nu\rho\sigma}\sqrt{g}d^4x.     \lb{32} \ee
The Hodge-star operation in the curved space-time determines the
following dual tensor:
\be   F^{*j\,\mu\nu} = \frac 12 \sqrt g
\epsilon^{\mu\nu\rho\sigma}F_{\rho\sigma}^j, \lb{33} \ee
and we have:
\be F^i\wedge F^j = \frac 12 F_{\mu\nu}^i F^{*j\,\mu\nu}d^4x\equiv
\frac 12 F^i\cdot F^{*\,j}d^4x. \lb{34} \ee
The requirement of self-duality in the curved space-time is absent
in gravity:
$$F^{*j\,\mu\nu}\neq F^{j\,\mu\nu}.$$
Then we obtain the following partition function:
\be
 Z = \int[{\cal D}A ][{\cal D}\psi] e^{-I}\approx \int
[{\cal D}A ][{\cal D}\psi]{\rm exp}\left[-\frac{1}{2}\int
(\Lambda\delta_{ij} + \psi_{ij})^{-1} F^i\cdot F^{*\,j}d^4x
\right]. \lb{35} \ee

In the limit of large $F^2\simeq 1$ ($M_{Pl}^{red}=1$ in our
units), the effective charge $g_{eff}$ of gravitational fields is
asymptotically small and equal to:
\be  g_{eff}^2\approx \frac{\Lambda}{2} \ll 1, \lb{40r} \ee
which follows from Eq.~(\ref{35}): minimal $g_{eff}$ corresponds to
$\psi_{ij}=0$. In such a regime (at small distances $r\sim
\lambda_{Pl}$) the space-time is Euclidean. This means that in our
(visible) Universe, we have respectively the flat Minkowski
space-time, and we can consider the condition of self-duality:
$F=F^*$.
Then the asymptotic gravitational Lagrangian is:
\be L_{eff}^{as} = - \frac{1}{4g_{eff}^2}F_{\mu\nu}^i F_{\mu\nu}^i
\approx - \frac{1}{2\Lambda}F_{\mu\nu}^i F_{\mu\nu}^i. \lb{41r}
\ee
However, at large distances the theory of gravity is more complicated
and the effective Lagrangian depends on $F^i\cdot F^{*\,i}$  (see
for example Refs.~\ct{30}).

\subsection{Coupling to matter}\label{sec6.1}

A complete theory of gravity can be constructed only if all matter
fields are incorporated into the theory. Then it is necessary to
construct the Lagrangian of the Universe considered in
Eq.~(\ref{71}).

In the Plebanski--Ashtekar formulation, the fundamental objects are a
rule for parallel transport with a connection in the curved space.
An operator which compares fields at different points is an
operator of the parallel transport between the points $x$ and $y$:
\be  U(x,y) = Pe^{i\int_{C_{xy}} \hat{A}_{\mu}(\tilde x)d{\tilde
x}^{\mu}},
   \lb{45r} \ee
where $P$ is the path ordering operator, $C_{xy}$ is a curve from
point $x$ till point $y$ and
 $$ \hat{A}_{\mu}(x) = A_{\mu}^i\frac{\sigma^i}{2},$$
with $\sigma^i$ being the Pauli matrices.\\
The operator:
\be  W = Pe^{i\oint \hat{A}_{\mu}(x)dx^{\mu}} \lb{46r} \ee
is the well-known Wilson loop. The graviton is related to a closed
string.

In the case of a spinor field $\chi(x)$ interacting with the gauge
field $A_{\mu}^i$, we have an additional gauge invariant
observable:
\be  \bar \chi(y)P e^{i\int_{C_{xy}} \hat{A}_{\mu}(\tilde
x)d{\tilde x}^{\mu}}\chi(x). \lb{47r} \ee
In the theory with two worlds, ordinary and mirror (or hidden),
the self-dual gravity with connection $A_{\mu}^i$ interacts with the
left-handed spinors $\chi_L$ and $\chi'_L$ of the visible and
mirror worlds, respectively, while the anti-self-dual gravity with
connection ${\bar A}_{\mu}^i$ interacts with the right-handed  spinors
$\chi_R$ and $\chi'_R$ of the O- and M-worlds, respectively.

Due to CP violation, the following cross-sections with ordinary
quarks $q$ and mirror quarks $q'$:
\be
\sigma(q + q \rightarrow q' + q') \neq \sigma(\bar q + \bar q
\rightarrow \bar q' + \bar q') \lb{47r} \ee
are different from each other, what is essential for the
baryogenesis.

\subsection{Quantum gravity and renormalization problem}\label{sec6.2}

In the framework of quantum field theory, and using the standard
techniques of perturbative calculations, one finds that
gravitation is non-renormalizable.

The theory of Loop Quantum Gravity (LQG) is a way of quantizing the Plebanski--Ashtekar gravity. In LQG, space is represented by a
spin network, evolving over time in discrete steps \ct{33,34}. The
phase space version \ct{31} of the new 'pure connection' viewpoint
on GR in the Plebanski formalism has led to the approach of LQG
\ct{34}. This class of theories is closed under the
renormalization \ct{30}.

In Refs. \ct{33} and \ct{34} it was argued that it is
possible to use Wilson loops as the basis for a nonperturbative
quantization of gravity. Explicit (spatial) diffeomorphism
invariance of the vacuum state plays an essential role in the
regularization of the Wilson loop states. An
explicit basis of states of quantum geometry was obtained, and the
geometry was shown to be quantized -- that is, the (non-gauge-invariant)
quantum operators representing area and volume have a discrete
spectrum. In this context, spin networks arose as a generalization
of Wilson loops. Plebanski's formalism is a starting point for
"spinfoam" models (see \ct{34} and references therein).

Should LQG succeed as a quantum theory of gravity, the known
matter fields will have to be incorporated into the theory.

Considering the problem of renormalizability of quantum gravity,
one can construct a model of multi-gravitons (see for example
\ct{35,36}) with $N$ massive gravitons.

\section{Summary and Conclusions}\label{sec7}

In this paper we have explained the main idea of Plebanski \ct{1}
to construct the 4-dimensional theory of gravity described by the
gravitational action with an integrand presented by a product of
two 2-forms, which are constructed from the tetrads $\theta^I$ and
the connection $A^{IJ}$ considered as independent dynamical
variables. Both $A^{IJ}$ and $\theta^I$ are 1-forms. The tetrads
$\theta^I_{\mu}$ were used instead of the metric $g_{\mu\nu}$. We
considered  the Minkowski space with the group of symmetry
$SO(1,3)$.

We have reviewed the well-known Plebanski BF-theory of general
relativity (GR) and constructed the gravitational actions of the
different theories of pure gravity: ordinary, dual and 'mirror'
ones, as well as the gravity with torsion. We have considered the
self-dual left-handed gravity of the Ordinary World (OW) and the
anti-self-dual right-handed gravity of the Mirror World (MW) with
broken mirror parity. We have shown that in the Plebanski
self-dual formulation of gravity the ordinary and dual
gravitational actions coincide.

We reviewed the close analogy of geometry of space-time in GR with
a structure of defects in a crystal \ct{21}. We have considered
the translational defects -- dislocations, and the rotational
defects -- disclinations, in the 4-dimensional crystals. The
crystalline defects represent a special version of the curved
space-time -- the Riemann--Cartan space-time with torsion \ct{7}.
The world crystal is a model for Einstein's gravitation which has
a new type of gauge symmetry with zero torsion as a special gauge,
while a zero connection (with zero Cartan curvature) is another
equivalent gauge with nonzero torsion which corresponds to the
Einstein's theory of "teleparallelism" \ct{22}. Here we showed
that in the Plebanski formulation, the phase of gravity with
torsion is equivalent to the ordinary or dual gravity, and we can
exclude torsion as a separate dynamical variable.

We have considered the equations of motion which follow from the
Plebanski action of gravity with the tetrads, self-dual connection
and auxiliary fields $\psi_{ij}$. The vacuum Einstein's equations
were obtained in the framework of the Plebanski theory of gravity.
Integrating out the tetrads we constructed the gravitational action
containing only the connection and the auxiliary fields. The
integration of the action over the auxiliary fields $\psi_{ij}$
leads to a new type of formulation of the gravitational theory
with a 'pure connection'. Here the diffeomorphism invariant gauge
theory of gravity is developed where the only dynamical field is
an $SU(2)$ spin connection \ct{29,30}. This theory is a completely
new perspective on GR.

We have calculated the partition function and the effective
Lagrangian of this 4-dimensional gravity. We have considered the
asymptotic limit of this theory: the large values $F^2\sim
M_{Pl}^4$, which correspond to small transPlanckian distances
$r\sim \lambda_{Pl}$, where $\lambda_{Pl}$ is the Planck length.
At these small distances, the connection fields $A_{\mu}^i$ exist
in the flat (Euclidean or Minkowski) space-time, and the effective
gravitational coupling constant is given by the cosmological constant
$\Lambda$: $g_{eff}=\Lambda/2$. At large distances we envisage a
more complicated theory of gravity. A complete theory of gravity
has to be constructed only with couplings to matter.

Finally, we recalled the role of Plebanski's formalism in the theory of Loop Quantum Gravity,
which is a way of quantizing the Plebanski--Ashtekar theory of
gravity.

\section*{Acknowledgments}
We are grateful to Masud Chaichian for useful discussions. The
support of the Academy of Finland under the Projects No. 136539 and
No.140886 is acknowledged.


\begin{thebibliography}{99}
\bibitem{1}
J.~Plebanski, J. Math. Phys. {\bf 18}, 2511 (1977).
\bibitem{2}
R.~Capovilla, T.~Jacobson, J.~Dell, and L.~Mason, Class. Quant.
Grav. {\bf 8}, 41 (1991).
\bibitem{4}
K.~Krasnov,  Gen. Rel. Grav. {\bf 43}, 1 (2011),
arXiv:0904.0423 [gr-qc]; Class. Quant. Grav. {\bf 26}, 055002
(2009), arXiv:0811.3147 [gr-qc]; Class. Quant. Grav. {\bf 25},
025001 (2008); Mod. Phys. Lett. A {\bf 22}, 3013 (2007),
arXiv:0711.0697 [gr-qc].
\bibitem{3}
M.P.~Reisenberger, gr-qc/9804061.
\bibitem{5}
Eyo Eyo Ita III, {\it CDJ formulation from the instanton representation
of Plebanski gravity}, arXiv:0911.0604 [gr-qc].
\bibitem{5a}
E.~Buffenoir, M.~Henneaux, K.~Noui and Ph.~Roche, Class. Quant.
Grav. {\bf 21}, 5203 (2004).
\bibitem{6}
F.~Tennie and M. N. R.~Wohlfarth, Phys. Rev. D {\bf 82}, 104052
(2010), arXiv:1009.5595 [gr-qc].
\bibitem{7}
E.~Cartan, Comt. Rend. Acad. Science {\bf 174}, 593 (1922).
\bibitem{8}
L.~ Freidel and K.~Krasnov, Class. Quant. Grav. {\bf 25}, 125018
(2008), arXiv:0708.1595 [gr-qc].
\bibitem{8a}
D.L.~Bennett, L.V.~Laperashvili, H.B.~Nielsen, A.~Tureanu,
arXiv:1206.3497 [gr-qc].
\bibitem{9}
A. Ashtekar, Phys. Rev. D {\bf 36}, 1587 (1987); Phys. Rev. Lett.
{\bf 57}, 1 (1986); {\it New Perspectives in Canonical Gravity},
(Bibliopolis, Napoli, 1988).
\bibitem{10}
C. R.~Das, L. V.~Laperashvili and A.~Tureanu, Eur. Phys. J. C {\bf 66},
307 (2010), arXiv:0902.4874 [hep-ph]; AIP Conf. Proc. {\bf 1241},
639 (2010), arXiv:0910.1669 [hep-ph]; Phys.Part.Nucl. {\bf 41},
965 (2010), arXiv:1012.0624 [hep-ph].
\bibitem{11}
C. R.~Das, L. V.~Laperashvili, H. B.~Nielsen and A.~Tureanu,
arXiv:1101.4558 [hep-ph]; Phys. Rev. D {\bf 84}, 063510 (2011),
arXiv:1105.6286 [hep-ph]; Phys. Lett. B {\bf 696}, 138 (2011),
arXiv:1010.2744 [hep-ph].
\bibitem{12}
T. D.~Lee and C. N.~Yang, Phys. Rev. {\bf 104}, 254 (1956).
\bibitem{13}
I. Yu.~Kobzarev, L. B.~Okun and I. Ya.~Pomeranchuk, Yad. Fiz. {\bf
3}, 1154 (1966) [Sov. J. Nucl. Phys. {\bf 3}, 837 (1966)].
\bibitem{13a}
K.~Nishijima and M.H.~Saffouri, Phys.Rev.Lett. {\bf 14}, 205
(1965).
\bibitem{13b}  E.~W.~Kolb, D.~Seckel, M. S.~Turner,
Nature {\bf 314}, 415 (1985); Fermilab-Pub-85/16-A, Jan. 1985.
\bibitem{14}
Z.~Berezhiani, A.~Dolgov and R. N.~Mohapatra, Phys. Lett. B {\bf
375}, 26 (1996), hep-ph/9511221; Z.~Berezhiani, {\it Through the
looking-glass: Alice's adventures in mirror world}, in: Ian Kogan
Memorial Collection ``From Fields to Strings: Circumnavigating
Theoretical Physics'', Eds. M.~Shifman et al., World Scientific,
Singapore, Vol.~3, pp. 2147-2195, 2005, hep-ph/0508233.
\bibitem{14a}
R.~Foot, H.~Lew and R. R.~Volkas, Phys. Lett. B {\bf 272}, 67
(1991); Mod. Phys. Lett. A {\bf 7},  2567 (1992); JHEP {\bf 0007},
032 (2000); R.~Foot, Mod. Phys. Lett. A {\bf 9}, 169 (1994);
Int. J. Mod. Phys. D {\bf 13}, 2161 (2004).
\bibitem{15}
Jian-Wei Cui, Hong-Jian He,  Lan-Chun Lu and Fu-Rong Yin,
arXiv:1110.6893 [hep-ph].
\bibitem{16}
A. G.~Riess et.~al., Astron.J. {\bf 116}, 1009 (1998),
astro-ph/9805201; S.~J.~Perlmutter et.~al., Nature {\bf 391}, 51
(1998), astro-ph/9712212; Astrophys. J. {\bf 517}, 565 (1999),
astro-ph/9812133; C.~L.~Bennett et~al., Astrophys. J. Suppl. {\bf
148}, 1 (2003), astro-ph/0302207; D.~N.~Spergel et~al., Astrophys.
J. Suppl. {\bf 148}, 175 (2003), astro-ph/0302209; Astrophys. J.
Suppl. {\bf 170}, 377 (2007), astro-ph/0603449; P.~Astier et ~al.,
Astron. Astrophys. {\bf 447}, 31 (2006), astro-ph/0510447; A.
Riess et al., Astrophys. J. Suppl. {\bf 183}, 109 (2009),
arXiv:0905.0697; W.~L.~Freedman et al., Astrophys. J. {\bf 704},
1036 (2009), arXiv:0907.4524.
\bibitem{15a}
S.~I.~Blinnikov and M.~Yu.~Khlopov, Sov. J. Nucl. Phys. {\bf 36},
472 (1982); Sov. Astron. {\bf 27}, 371 (1983); H.~M.~Hodges, Phys.
Rev. D {\bf 47}, 456 (1993).
\bibitem{16a}
H.~An, S.~L.~Chen, R.~N.~Mohapatra and Y.~Zhang, JHEP 03, 124
(2010), arXiv:0911.4463.
\bibitem{16b}
L.~B.~Okun, Phys. Usp. {\bf 50}, 380 (2007), hep-ph/0606202;
S.~I.~Blinnikov, {\it Notes on Hidden Mirror World},
arXiv:0904.3609; P.~Ciarcelluti, Int. J. Mod. Phys. D {\bf 19},
2151 (2010), arXiv:1102.5530, and references therein.
\bibitem{16c}
L.~Bento and Z.~Berezhiani, Phys. Rev. Lett. {\bf 87}, 231304 (2001)
[hep-ph/0111116];  Fortsch. Phys. {\bf 50}, 489 (2002).
\bibitem{16d}
M.~Fukugita and T.~Yanagida, Phys.Lett. B {\bf 174}, 45 (1986);
W.~Buchmuller, R.~D.~Peccei and T.~Yanagida, Ann. Rev. Nucl. Part. Sci.
{\bf 55}, 311 (2005), arXiv:hep-ph/0502169.
\bibitem{20a}
L. B.~Okun, JETP Lett. {\bf 31}, 144 (1980); Pisma Zh. Eksp.Teor.
Fiz. {\bf 31}, 156 (1979); Nucl. Phys. B {\bf 173}, 1 (1980).
\bibitem{20b}
R. N.~Mohapatra and V. L.~Teplitz, Phys. Lett. B {\bf 462}, 302
(1999).
\bibitem{20c}
S. L.~Glashow, Phys.Lett. B {\bf 167}, 35 (1986); A.~Badertscher
et al., Int. J. Mod. Phys. A {\bf 19}, 3833 (2004).
\bibitem{20d}
V.~Berezinsky and A.~Vilenkin, Phys. Rev. D {\bf 62}, 083512
(2000); Z.~K.~Silagadze, Phys. Atom. Nucl. {\bf 60}, 272 (1997)
[Yad. Fiz. {\bf 60}, 336 (1997)].
\bibitem{20e}
R.~N.~Mohapatra, S.~Nasri and S.~Nussinov, Phys. Lett. B {\bf
627}, 124 (2005); Z.~Berezhiani and L.~Bento, Phys. Rev. Lett.
{\bf 96}, 081801 (2006); Phys. Lett. B {\bf 635}, 253 (2006); Yu.~
N.~Pokotilovski, Phys. Lett. B {\bf 639}, 214 (2006).
\bibitem{20f}
A.~Yu.~Ignatiev and R.~R.~Volkas, Phys. Lett. B {\bf 487}, 294
(2000).
\bibitem{21}
H.~Kleinert,  {\it Multivalued Fields in Condensed Matter,
Electromagnetism, and Gravitation} (World Scientific, Singapore,
2008); EJTP {\bf 7}, 287 (2010).
\bibitem{21a}
D.~L.~Bennett and H.~B.~Nielsen, in: Proceedings of the 14th
Workshop "What Comes Beyond the Standard Model", Bled, Slovenia,
July 11-21, 2011.
\bibitem{21b}
D~L.~Bennett, C~R.~Das, L.~V.~Laperashvili, H.~B.~Nielsen,
arXiv:1209.2155 [hep-th].
\bibitem{22}
 E.~Cartan and A.~Einstein, {\it Letters of Absolute Parallelism} (Princeton University
Press, Princeton, NJ, 1922).
\bibitem{26}
Z.~Bern, {\it Perturbative quantum gravity and its relation to
gauge theory}, Living Rev. Rel. {\bf 5}, 5 (2002), gr-qc/0206071.
\bibitem{27}
Z.~Bern, T.~Dennen, Y.~T.~Huang and M.~Kiermaier, Phys. Rev. D
{\bf 82}, 065003 (2010), arXiv:1004.0693 [hep-th].
\bibitem{28}
T.~Jacobson and L.~Smolin, Class. Quant. Grav. {\bf 5}, 583 (1988).
\bibitem{29}
R.~Capovilla, T.~Jacobson and J.~Dell, Phys. Rev. Lett. {\bf 63},
2325 (1989); Class. Quant. Grav. {\bf 8}, 59 (1991).
\bibitem{30}
K.~Krasnov, Phys. Rev. D {\bf 84}, 024034 (2011), arXiv:1101.4788;
Phys. Rev. Lett. {\bf 106}, 251103 (2011), arXiv:1103.4498.
\bibitem{31}
I.~Bengtsson, Phys. Lett. B {\bf 254}, 55 (1991).
\bibitem{32}
K.~Krasnov, Phys. Rev. D {\bf 81}, 084026 (2010), arXiv:0911.4903
[hep-th]; Europhys. Lett. {\bf 89}, 30002 (2010), arXiv:0910.4028
[gr-qc].
\bibitem{33}
L.~Smolin, {\it Three Roads to Quantum Gravity} (Oxford University
Press, 2000).
\bibitem{34}
C.~Rovelli, {\it Quantum Gravity} (Cambridge Monographs on
Mathematical Physics, 2003); {\it Loop quantum gravity}, Living
Rev. Rel. {\bf 11}, 5 (2008).
\bibitem{35}
Hajime Isimori, arXiv:1010.5122 [gr-qc].
\bibitem{36}
L.~V.~Laperashvili, arXiv:1107.5927 [gr-qc].

\end{thebibliography}
\end{document}